\documentclass[aps,prb,preprint,groupedaddress]{revtex4-1}
\usepackage{graphicx}
\usepackage{amsmath}
\usepackage{amsfonts}
\usepackage{textcomp}

\begin{document}
\newcommand{\guido}[1]{[\textbf{GUIDO:} \emph{#1}]}
\newcommand{\miquel}[1]{[\textbf{MIQUEL:} \emph{#1}]}

\title{Prediction of inelastic light scattering spectra from electronic collective excitations in GaAs/AlGaAs core-multishell nanowires}

\author{Miquel Royo$^{1,2}$}
\email{mroyo@qfa.uji.es}
\author{Andrea Bertoni$^{2}$}
\author{Guido Goldoni$^{2,3}$}

\affiliation{$^{1}$Departament de Qu\'{\i}mica F\'{\i}sica i Anal\'{\i}tica, Universitat Jaume I, E-12080, Castell\'o, Spain}
\affiliation{$^{2}$CNR-NANO S3, Institute for Nanoscience, Via Campi 213/a, 41125 Modena, Italy}
\affiliation{$^{3}$Department of Physics, Informatics and Mathematics, University of Modena and Reggio Emilia, Via Campi 213/a, 41125 Modena, Italy}

\date{\today}

\begin{abstract}
We predict inelastic light scattering spectra from electron collective excitations in a coaxial quantum well embedded in a core-multishell $\mathrm{GaAs/AlGaAs}$ nanowire. The complex composition, the hexagonal cross section and the remote doping of typical samples are explicitly included, and the free electron gas is obtained by a DFT approach. Inelastic light scattering cross sections due to charge and spin collective excitations belonging to quasi-1D and quasi-2D states, which coexist in such radial heterostructures, are predicted in the non-resonant approximation from a fully three-dimensional multi-subband TDDFT formalism. We show that collective excitations can be classified in azimuthal, radial and longitudinal excitations, according to the associated density fluctuations, and we suggest that their character can be exposed by specific spectral dispersion of inelastic light scattering  along different planes of the heterostructure.

\end{abstract}

\pacs{}

\maketitle


\section{Introduction \label{intro}}

Inelastic light scattering (ILS) spectroscopy is a fundamental optical technique to study electronic properties of excess carriers in semiconductors.~\cite{ILScardonaBook,PinczukBook89,SchullerBook2006,KushwahaSSR2011} First, it enables to probe electronic elementary excitations, which in semiconductors occur in the range of few tens of meV, with near-infrared and visible light for which high-quality tunable lasers and detectors exist. Second, ILS enables to study separately different types of elementary excitations thanks to the polarization selection rules.~\cite{SchullerBook2006} 
Indeed, the ILS cross section arises from collective charge density excitations (CDEs), or plasmons, when the polarizations of the incident and scattered light are parallel (polarized configuration), and from collective spin density excitations (SDEs) when the two polarizations are orthogonal (depolarized configuration). In addition, single-particle excitations (SPEs) can be observed in both polarization configurations under strong inter-band resonance conditions.~\cite{SchullerPRB96} 
Finally, by adjusting the relative direction of the incident and scattered photons, it is possible to tune the momentum transferred to the electronic system in a given direction, which, due to the momentum dependent selection rules,~\cite{SchullerBook2006} enables to map the dispersion of the elementary excitations. 

The wavevector dispersion of elementary excitations is strongly dependent on the dimensionality of the electronic system.~\cite{DasSarmaPRB96,KongPST07,KushwahaSSR2011} This is particularly important for intra-subband plasmons, as well as the Landau damping in SPEs continua. Such features have been demonstrated in ILS experiments in semiconductor heterostructures of reduced dimensionality such as quantum wells (QWs),~\cite{BlumPRB70,KatayamaJPSJ85,HawrylakPRB85,DasSarmaPRL99, KushwahaAIPA12} quantum wires~\cite{SteinebachPRB96,TavaresPRB05,DasSarmaPRB96} and quantum dots.~\cite{Garcia2005,Kalliakos2008} The comparison of the experimental results with theoretical models yielded information about the carriers energy structure, density, mobility and many-body contributions.

Recently, ILS experiments and theoretical predictions have been used to confirm the accumulation of excess electrons in the conduction band of radial heterostructures in nanowires, namely, core-multishell nanowires (CSNW) hosting a coaxial quantum well (coQW), similar to the one shown in Fig.~\ref{fig1}(a).~\cite{FunkNL13} The extra signals observed in the ILS spectra of remotely-doped coQWs, in comparison with those of undoped reference samples, were shown to originate from the collective excitations of a high-mobility electron gas confined inside the coQW. Furthermore, ILS resonances could be selectively assigned to quasi-one-dimensional (q1D) and quasi-two-dimensional (q2D) states which coexist in the structure, due to the hexagonal nanowire cross section, being localized at the corners and facets of the coQW, respectively. Such a spatial-dependent dimensionality and the tubular shape of the electron channel, which bridges between q2D and a q1D electron gas, guarantee a particular dispersion of the collective excitations in the confined in-plane directions and a diameter-dependent momentum dispersion along the invariant longitudinal direction. However, results in Ref.~\onlinecite{FunkNL13} were restricted to a backscattering geometry with incidence normal to the facets of the coQW and did not probe the energy dispersion of such excitations.

It is important to note that ILS in coQWs substantially differs from ILS in planar structures. On the one hand, in coQWs there is only one translational invariant direction (along the nanowire axis), the other direction being wrapped around the CSNW core, removing the continuous energy dispersion. The discrete symmetry also induces a inhomogeneous carrier distribution, with coexisting q1D and q2D states (see Fig.~\ref{fig1}(b)). On the other hand, the photon field always impinges on more than one coQW planes, and with different angles, at the same time (see Fig.~\ref{fig1}(a)). Since selection rules depend on the photon field direction, ILS spectra in coQWs are expected to be substantially more complex than in traditional QWs.

The dispersion of plasmon modes has been theoretically investigated in cylindrical tubular geometries.~\cite{LinPRB93,SatoPRB93,WendlerPRB94,TanatarPRB97,WangPRA02} These studies, mostly RPA calculations, illustrated the dimensional crossover from q2D to q1D electron gas for very small radius of the tubule, a regime appropriate, e.g., to carbon nanotubes but not to CSNWs. Furthermore, these works neglected any discrete symmetry characteristic of the nanowire cross section. The lowest collective excitations of an electron gas in CSNWs, including the hexagonal symmetry of the electronic system, have been studied recently by our group,~\cite{RoyoPRB14} which enabled to rationalize the complex excitation spectra in terms of symmetry allowed excitations and symmetry selective Landau damping. However, the high electron density regime, relevant for ILS experiments, and the ensuing self-consistent field was not included. Likewise, ILS dispersion was not discussed.

In this paper, we study theoretically the ILS spectra of a prototypical $\mathrm{GaAs/Al_{0.3}Ga_{0.7}As}$ CSNW hosting a remotely doped coQW at large electron densities.~\cite{KettererPRB11,FunkNL13} Assuming non-resonant conditions, the scattering cross section is obtained from the dynamic response functions of the excess electrons. Calculations of CDE and SDE spectra are performed in a 3D real-space multi-subband DFT and TDDFT formalism which accounts for the complex geometry and composition of the structure, the self-consistent field of excess electrons, and the plasmon-phonon coupling. Three different classes of collective excitations arise in the spectra, which involve density fluctuations occurring in the azimuthal, radial or longitudinal directions of the coQW. We show that these collective modes can be singled out by properly designed ILS experiments.

\section{Theoretical model \label{theory}}

\subsection{Self-consistent ground-state calculation \label{gscalc}}

To obtain the electronic states of a CSNW we employ a standard envelope function approach in a
single parabolic band approximation. Electron-electron interactions have been treated at the mean-field level through a typical Kohn-Sham (KS) LDA procedure. We consider a CSNW which is spatially invariant along the NW axis direction z (see Fig.~\ref{fig1}(a) for axis definitions). Therefore, we factorize the envelope functions as $\Psi_{n,k_z}(\mathbf{r},z)=\varphi_{n}(\mathbf{r})\,e^{ik_zz} $, with parabolic energy dispersion $\varepsilon_n(k_z)=\epsilon_n + \frac{\hbar^2\,k_z^2}{2\,m^*_e} $. In the $\mathbf{r}\equiv (x,y)$ plane, the NW cross section is hexagonal 
and the material and doping modulations are mapped on a hexagonal domain using a 
symmetry-preserving triangular grid. The self-consistent potential $v_{KS}(\boldsymbol{r})=v(\boldsymbol{r})+v^H(\boldsymbol{r}) + v^{XC}(\boldsymbol{r})$ includes the effect of the spatial confinement $v(\boldsymbol{r})$ determined by the materials conduction band offset, the Hartree potential $v^H(\boldsymbol{r})$ generated by the free electrons and the static doping, and an approximate exchange-correlation potential $v^{XC}(\boldsymbol{r})$.~\cite{GunnarsonPRB76} Further details can be found in Ref.~\onlinecite{BertoniPRB2011}. 

\subsection{ILS spectra calculation \label{ILScalc}}

ILS spectra have been calculated within a multi-subband TDLDA non-resonant formalism.~\footnote{A thorough description of the computational approach can be found in Ref.~\onlinecite{RoyoPRB14}.}
The scattering cross section due to CDEs is obtained from the imaginary part of the momentum-dependent density-density response function (DDRF),~\cite{KushwahaAIPA12,KushwahaAIPA13} 

\begin{equation}
I_{CDE}(\mathbf{Q},\omega)\propto -\Im\left[\tilde{\Pi}(\mathbf{Q}, \omega)\right] = -\Im\left[  
\iint d\mathbf{r} \, d\mathbf{r}' e^{-i\,\mathbf{q}(\mathbf{r}-\mathbf{r}')} \tilde{\Pi}(\mathbf{r},\mathbf{r'},q_z,\omega) \right],
\label{eq1}
\end{equation}

\noindent where, $\mathbf{q}$ is the in-plane and $q_z$ the in-wire, longitudinal components of the total momentum $\mathbf{Q}\equiv(\mathbf{q},q_z)$ exchanged in the scattering process, i.e., $\mathbf{Q}=\mathbf{Q}_i-\mathbf{Q}_s$, $\mathbf{Q}_i$ and $\mathbf{Q}_s$ being the momenta of the incident and scattered photons, respectively, and $\omega=\omega_i-\omega_s$ is the ILS energy shift. To obtain the dynamic DDRF, we expand $\tilde{\Pi}(\boldsymbol{r},\boldsymbol{r}',q_z,\omega)$ in terms of the in-plane KS envelope functions,

\begin{equation}
\tilde{\Pi}(\boldsymbol{r},\boldsymbol{r}',q_z,\omega)=\sum_{ijlm} \tilde{\Pi}_{ijlm}(q_z,\omega) \,
\varphi^*_i(\boldsymbol{r})\, \varphi_j(\boldsymbol{r})\,\varphi_l(\boldsymbol{r}')\,\varphi^*_m(\boldsymbol{r}'),
\label{eq2}
\end{equation}

\noindent where the elements $\tilde{\Pi}_{ijlm}(q_z,\omega)$ are defined by the Dyson equation

\begin{equation}
\sum_{ijlm}\tilde{\Pi}_{ijlm}(q_z,\omega)=
\sum_{ijlm}  \Pi^0_{ij}(q_z,\omega)\, \delta_{il}\,\delta_{jm} +
\sum_{ij} \Pi^0_{ij}(q_z,\omega)
\sum_{knlm}
\left[v_{ijkn}(q_z)+u^{XC}_{ijkn} \right]
\tilde{\Pi}_{knlm}(q_z,\omega).
\label{eq3}
\end{equation}

\noindent Here, $\Pi^0_{ij}(q_z,\omega)$ are the elements of the DDRF for the KS system,

\begin{equation}
\Pi_{ij}^0(q_z,\omega)= g \int \frac{dk_z}{2\pi} \frac{f_i(k_z)-f_j(k_z+q_z)}{\omega-(\varepsilon_j(k_z+q_z)
-\varepsilon_i(k_z))+i\eta},
\label{eq4}
\end{equation}

\noindent where $g=2$ accounts for electron spin degeneracy, $f_n(k_z)$ is the Fermi occupation function and $\eta$ is the electron damping parameter. $v_{ijkn}(q_z)$ and $u^{XC}_{ijkn}$ are the direct Coulomb and exchange-correlation matrix elements, respectively, which describe the dynamic interaction of two electrons, one of which gets scattered from state $i$ to $j$ and the other from $k$ to $n$, with an exchange of momentum $q_z$.

The direct Coulomb matrix elements read

\begin{equation}
v_{ijkn}(q_z)= \int d\boldsymbol{r}
\int d\boldsymbol{r}' \, \phi_i(\boldsymbol{r})\, \phi_j^*(\boldsymbol{r}) \,
\hat{V}(\boldsymbol{r}-\boldsymbol{r}',q_z)\,
\phi^*_k(\boldsymbol{r}')\, \phi_n(\boldsymbol{r}'),
\label{eq5}
\end{equation}

\noindent where $\hat{V}(\boldsymbol{r}-\boldsymbol{r}',q_z)$ is the Fourier transform of the Coulomb operator in the
$z$ direction. We have used a frequency dependent dielectric constant $\varepsilon(\omega)$, entering the denominator of the Coulomb operator instead of the semiconductor high-frequency dielectric constant $\varepsilon_{\infty}$, that phenomenologically accounts for the coupling of CDEs with the phonons of the underlying crystal lattice,

\begin{equation}
\varepsilon(\omega)=\varepsilon_{\infty}\frac{\omega^2-\omega_{LO}^2+i\,\Gamma_{ph}\,\omega}
{\omega^2-\omega_{TO}^2+i\,\Gamma_{ph}\,\omega}.
\label{eq6}
\end{equation}

\noindent Here, $\omega_{LO}$ and $\omega_{TO}$ are the resonant frequencies of the longitudinal and transverse phonons, respectively, and $\Gamma_{ph}$ is a damping parameter.

The exchange-correlation matrix elements read 

\begin{equation}
u^{XC}_{ijkn}= - \int d\boldsymbol{r} \int d\boldsymbol{r}'
 \, \phi_i(\boldsymbol{r})\, \phi_j^*(\boldsymbol{r}) \,
\hat{f}_{XC}(\boldsymbol{r},\boldsymbol{r}')\,
\phi^*_k(\boldsymbol{r}')\, \phi_n(\boldsymbol{r}'),
\label{eq7}
\end{equation}

\noindent with $\hat{f}_{XC}(\boldsymbol{r},\boldsymbol{r}')$ being the dynamic exchange-correlation kernel defined in the adiabatic LDA as the derivative of the static exchange-correlation potential with respect to the ground state density, $\hat{f}_{XC}(\boldsymbol{r},\boldsymbol{r}') =\frac{\delta v^{XC} \left[n_0\right](\boldsymbol{r})} {\delta n_0(\boldsymbol{r}')} \delta(\boldsymbol{r}-\boldsymbol{r}')$. 

The TDLDA formalism also provides the ILS intensity due to SDEs. These are only coupled through indirect-exchange interactions. Consequently, they always appear redshifted with respect to their plasmonic counterparts. The ILS cross section due to SDEs is therefore obtained from the imaginary part of the so-called  momentum-dependent irreducible response function (IRF), $\Pi(\mathbf{Q},\omega)$, which is calculated as above for the DDRF, but with the direct Coulomb integrals set to zero in Eq.~(\ref{eq3}).

From the DDRF and the IRF we can obtain the density fluctuation induced by the electromagnetic field at a given
energy and momentum, the so-called induced density distribution (IDD), from Kubo's correlation formula.~\cite{KuboJPSJ57} Thus, for CDEs the IDD is given by

\begin{equation}
\delta n_{CDE}(\mathbf{r},\mathbf{q},q_z,\omega)= \int d\mathbf{r}' \, \tilde{\Pi}(\boldsymbol{r},\boldsymbol{r}',q_z,\omega)
\, e^{i\, \mathbf{q} \mathbf{r}'},
\label{eq8}
\end{equation}

\noindent whereas for SDEs, $\delta n_{SDE}(\mathbf{r},\mathbf{q},q_z,\omega)$ is equivalently calculated substituting $\tilde{\Pi}(\boldsymbol{r},\boldsymbol{r}',q_z,\omega)$ by $\Pi(\boldsymbol{r},\boldsymbol{r}',q_z,\omega)$.

\section{Numerical Results \label{Results}}

\subsection{Electron gas ground-state distribution \label{EGdist}}

\begin{figure}[ht!]
 \includegraphics[width=0.4\textwidth]{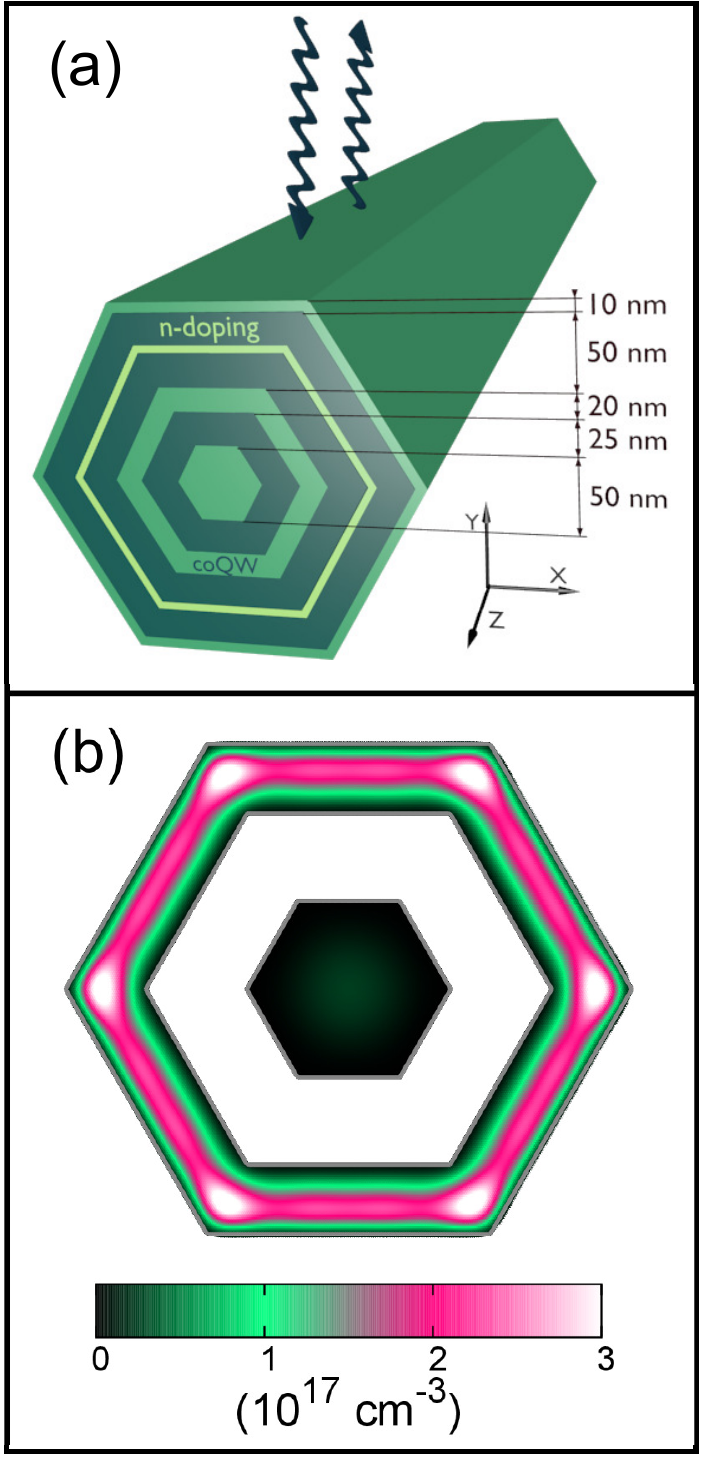}
\caption{(a) Schematics of a prototypical CSNW investigated in the paper. Dark, middle and light green colors are used for $\mathrm{Al_{0.3}Ga_{0.7}As}$, GaAs, and $n$-doping regions, respectively. (b) Calculated electron gas distribution over the CSNW cross section. Only the GaAs core and coQW regions are shown.}
\label{fig1}
\end{figure}

Calculations are performed for the CSNW shown in Fig.~\ref{fig1}(a). A GaAs core is surrounded by   $\mathrm{Al_{0.3}Ga_{0.7}As}$/GaAs/$\mathrm{Al_{0.3}Ga_{0.7}As}$/GaAs layers with thickness indicated in the figure. 
A $n$-type $\mathrm{\delta}$-doping layer is included in the middle of the outer $\mathrm{Al_{0.3}Ga_{0.7}As}$ layer. System parameters are similar to those of the sample recently studied in Ref.~\onlinecite{FunkNL13}, although the thickness of the GaAs coQW is slightly narrower here (26 nm in the experiment) to spectrally separate different classes of transitions, which are partially overlapping in Ref.~\onlinecite{FunkNL13}, as discussed in Sec.~\ref{Fixgeom}.

Material parameters for GaAs ($\mathrm{Al_{0.3}Ga_{0.7}As}$) are the isotropic electron effective mass $m^*_e=0.067 \, (0.092)$ and the static dielectric constant $\varepsilon=13.18\, (12.24)$. The GaAs/$\mathrm{Al_{0.3}Ga_{0.7}As}$ conduction band offset is taken as 284 meV. The Fermi level is pinned at the middle of the GaAs band gap (taken as 1.43 eV), the temperature is 4 K, and a common homogeneous triangular grid with $\sim3.21$ $\mathrm{nodes/nm^2}$ is used to numerically integrate the KS and Poisson equations. Unless otherwise stated, we have used a constant density of static donors over the $\mathrm{\delta}$-doping layer of $2\times10^{18}\, \mathrm{cm^{-3}}$, which yields an accumulated electron gas with a total density $\sim1.13\times10^{7}\, \mathrm{cm^{-1}}$.  As shown in Fig.~\ref{fig1}(b), electrons are almost exclusively localized in the GaAs coQW with only a minority of them distributed over the GaAs core. In the coQW, in turn, electrons are preferentially localized in the hexagonal corners in order to minimize electron-electron interactions and single-particle confinement energies. This is an electron distribution previously observed in different hexagonal CSNW heterostructures at sufficiently large electron density,~\cite{FerrariNL08,BertoniPRB2011,WongNL2011,RoyoPRB13,FunkNL13,JadczakNL14,RoyoPRB15} which points to coexisting channels of different dimensionalities, namely, q1D states in the corners and q2D states in the facets of the coQW.

\subsection{ILS at normal incidence \label{Fixgeom}}

We now study the ILS spectra due to the elementary excitations of the conduction free electrons accumulated in the CSNW in a back-scattering geometry (i.e., with antiparallel incident and scattered photons) and with the photon field at normal incidence with respect to the top facet of the coQW. In this configuration no momentum is transferred to the electronic system along the invariant CSNW axis direction. In the calculation, we have assumed a typical excitation energy of 1.92 eV. We approximate $\lvert\mathbf{Q_i}\rvert=\lvert\mathbf{Q_s}\rvert$, yielding a transferred momentum $\lvert\mathbf{Q}\rvert=1.38\times 10^6 \, \mathrm{cm^{-1}}$, which in our sample corresponds to $\sim 1.03 \, k_F$, $k_F$ being the Fermi wave-vector. We have used a basis set formed by the 100 lowest-lying KS subbands which assures convergence within the considered energy range. Finally, in order to account for the plasmon-phonon coupling effect we have used the GaAs high-frequency dielectric constant $\varepsilon_{\infty}=10.86$, the GaAs phonon resonances $\omega_{TO(LO)}=33.72\,(36.69)$ meV, and a damping parameter $\Gamma_{ph}=1$ meV. 

\begin{figure}[ht!]
\includegraphics[width=0.8\textwidth]{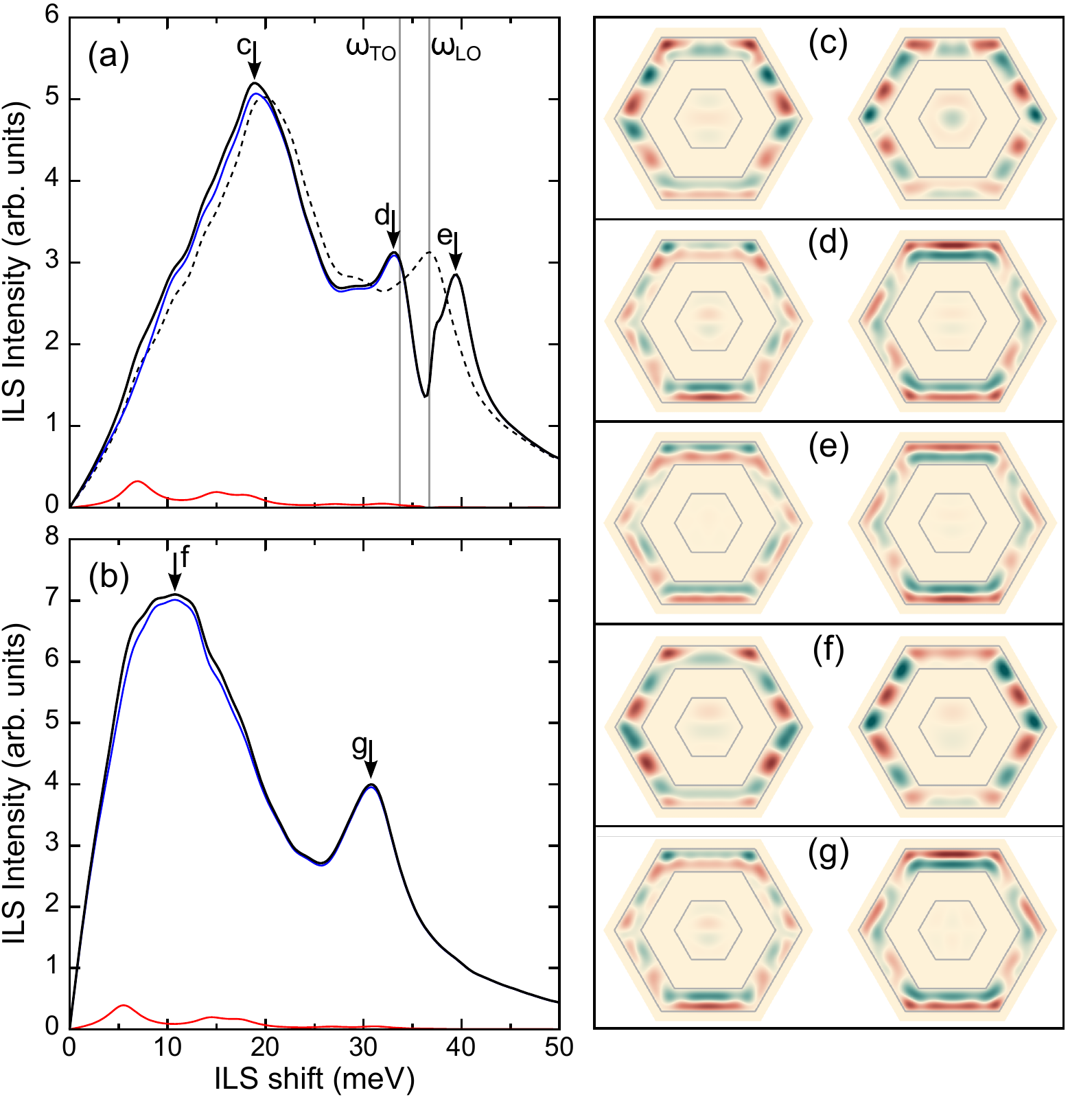}
\caption{CDE (a) and SDE (b) spectra in a back-scattering configuration with the photons at normal incidence on the top CSNW facet. Blue and red lines show the contribution to the total spectra from excitations occurring in the GaAs coQW and core, respectively. The dashed line in (a) shows the CDE spectrum calculated neglecting the plasmon-phonon coupling. Vertical gray lines mark the position of the GaAs LO and TO phonon resonances. (c)-(g) Real (left) and imaginary (right) parts of the IDDs calculated at the resonances indicated by arrows in the spectra (a) and (b).}
\label{fig2}
\end{figure}

In Fig.~\ref{fig2} we show the calculated CDE (a) and SDE (b) ILS spectra with an electron damping $\eta = 2$ meV, which leads to peak bandwidths comparable to those measured in recent ILS experiments in CSNWs.~\cite{FunkNL13} We also show a breakdown of the spectra in contributions due to excitations in the GaAs coQW (blue) and core (red). As can be seen, the cross section is almost entirely originated in coQW excitations, with some weak core excitations at low energies that hardly influence the total spectrum. Likewise, we have found no appreciable coupling between core and coQW excitations by comparing the spectra in Fig.~\ref{fig2} with equivalent ones obtained by deactivating the dynamic Coulomb and exchange-correlation integrals between core and coQW states (not shown). Such a result contrast with the strong inter-tubule coupling observed in coaxial tubules~\cite{LinPRB93,TanatarPRB97} in which, however, the inter-tubule separation was of the order of few \AA{}.

Both CDE and SDE spectra are dominated by a low-energy broad band, labeled \textsf{c} and \textsf{f} in Fig.~\ref{fig2}(a) and (b), peaking at $\sim18.9$ and $\sim10.8$ meV, respectively. From the energy of the corresponding peak in the SPE spectrum (not shown), we estimate a depolarization shift $\sim 5.8$ meV and an excitonic correction $\sim 2.8$ meV. As reported in Ref.~\onlinecite{FunkNL13}, this broad signal is due to excitations between the discrete subbands which arise from the discretization of the 2D continuum of an equivalent flat QW due to azimuthal periodicity of the coQW. Note that in a planar QW these collective modes would correspond to genuine intra-subband excitations and would not be excited at normal incidence. In our CSNW these modes are activated through the lateral facets which offer oblique orientation to the incident photons. By this reason, and in order to avoid confusion with the genuine 1D intra-subband excitations along the CSNW axis, we will refer to these excitations as \emph{2D intra-subband excitations}. 

This assignment is confirmed by the calculated IDDs shown in panels (c) and (f) of Fig.~\ref{fig2}, respectively. These clearly correspond to density fluctuations running along the lateral facets of the coQW with nodal surfaces \textit{normal} to the coQW planes (azimuthal nodes). No nodal surfaces \textit{parallel} to the QW plane (radial nodes) are present, since these are associated to inter-subband excitations lying at larger energies, as we discuss next.

The CDE spectrum in Fig.~\ref{fig2}(a) also shows two less intense peaks below and above the GaAs phonon resonances (gray vertical lines), labeled \textsf{d} and \textsf{e}. The spectrum calculated neglecting the plasmon-phonon coupling (dashed line in Fig.~\ref{fig2}(a)) demonstrates that these two peaks are the split pair of a single CDE resonance due to coupling with the phonons. The corresponding SDE peak, labeled \textsf{g} in Fig.~\ref{fig2}(b), is at $\sim 31$ meV.
This collective mode corresponds to the first inter-subband excitation in the coQW, as clearly demonstrated by the calculated IDDs shown in Figs.~\ref{fig2}(d),(e) and (g) featuring one radial node in the top and bottom facets. 
The depolarization shift for this mode (calculated using the energy of the CDE without phonon coupling) is $\sim 3.9$ meV and the excitonic correction is $\sim 2$ meV. Therefore, many-electron corrections are smaller for the inter-subband than for the 2D intra-subband mode, which is reasonable due to the higher confinement regime of the former. 
Note that the position of the inter-subband peaks is very sensitive to the coQW thickness, and for thicker coQWs (as in Ref.~\onlinecite{FunkNL13}) these modes are likely to merge to the broad 2D intra-subband peak.

\begin{figure}[ht!]
\includegraphics[width=0.4\textwidth]{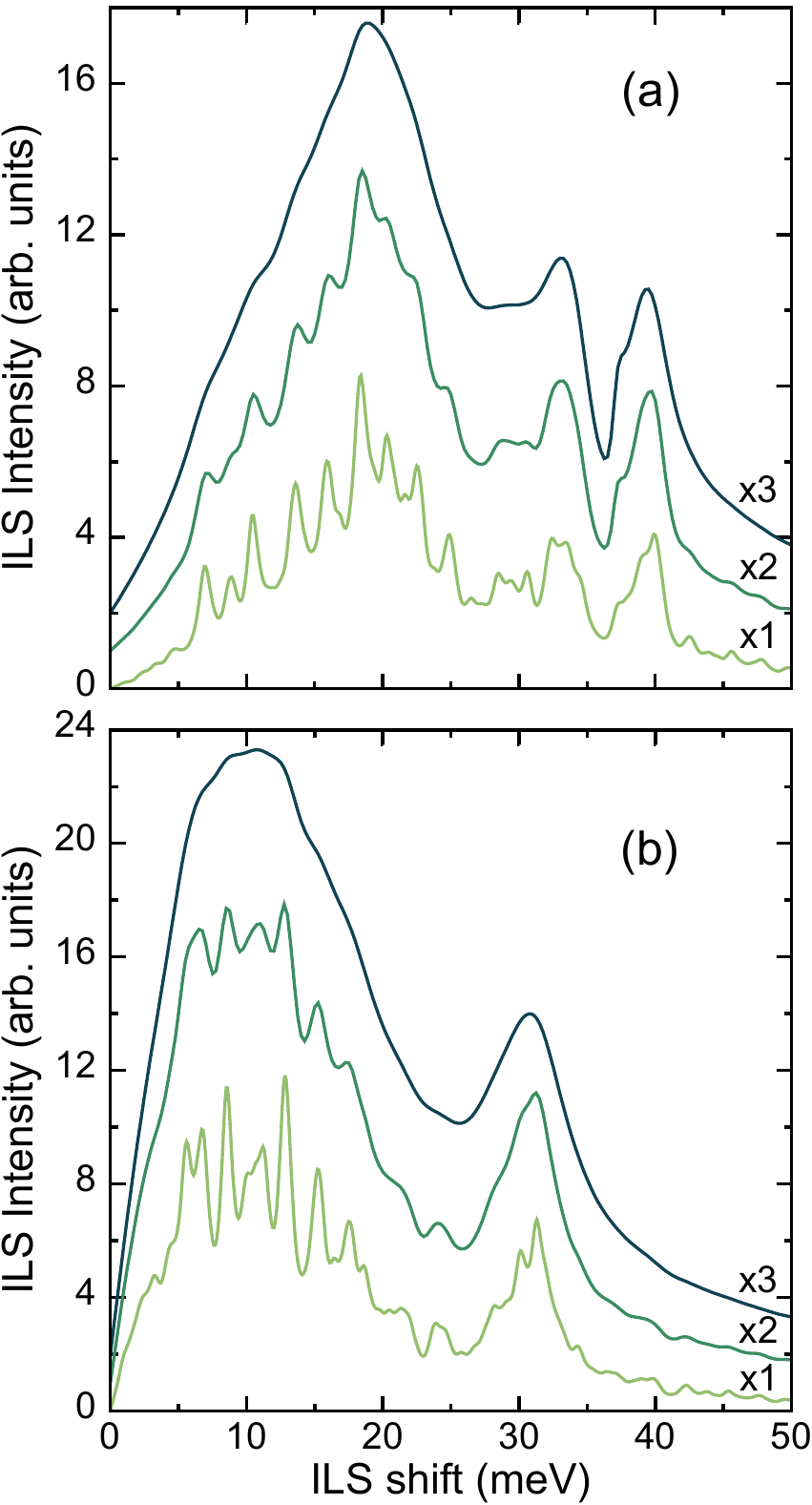}
\caption{(a) CDE and (b) SDE spectra calculated using an electron damping of 2, 1 and 0.5 meV from top to bottom lines. For clarity, the curves have been offset and scaled, as indicated by labels.}
\label{fig4}
\end{figure}

In Fig.~\ref{fig4} we show the effect of reducing the value of the electron damping, which might be achieved in samples with higher electron mobilities, on the calculated ILS spectra. In contrast to a flat QW, where the damping only modifies the bandwidth of single intra- or inter-subband excitations, in a CSNW a small damping exposes a complex fine-structure. Indeed, at the energy of the broad 2D intra-subband band, numerous close by collective excitations show up. As discussed above, these excitations are associated to transitions between states with different number of azimuthal nodes, arising from the azimuthal discretization of the 2D continuum. Also the inter-subband peak, which is mainly associated to transitions with a change in one radial node, shows several resonances at small damping, which correspond to different changes in the number of azimuthal nodes, as shown by the corresponding IDDs (Figs.~\ref{fig2}(d),(e) and (g)). Note indeed that the two types of excitations cannot be excited individually due to the different impinging angle on different facets. Since in Fig.~\ref{fig4} the fine structure emerges already at a damping of $\sim 1$ meV, the fine structure is likely to be observed in future ILS experiments. 

\subsection{Longitudinal dispersion  \label{Vargeom}}

\begin{figure}[ht!]
\includegraphics[width=0.4\textwidth]{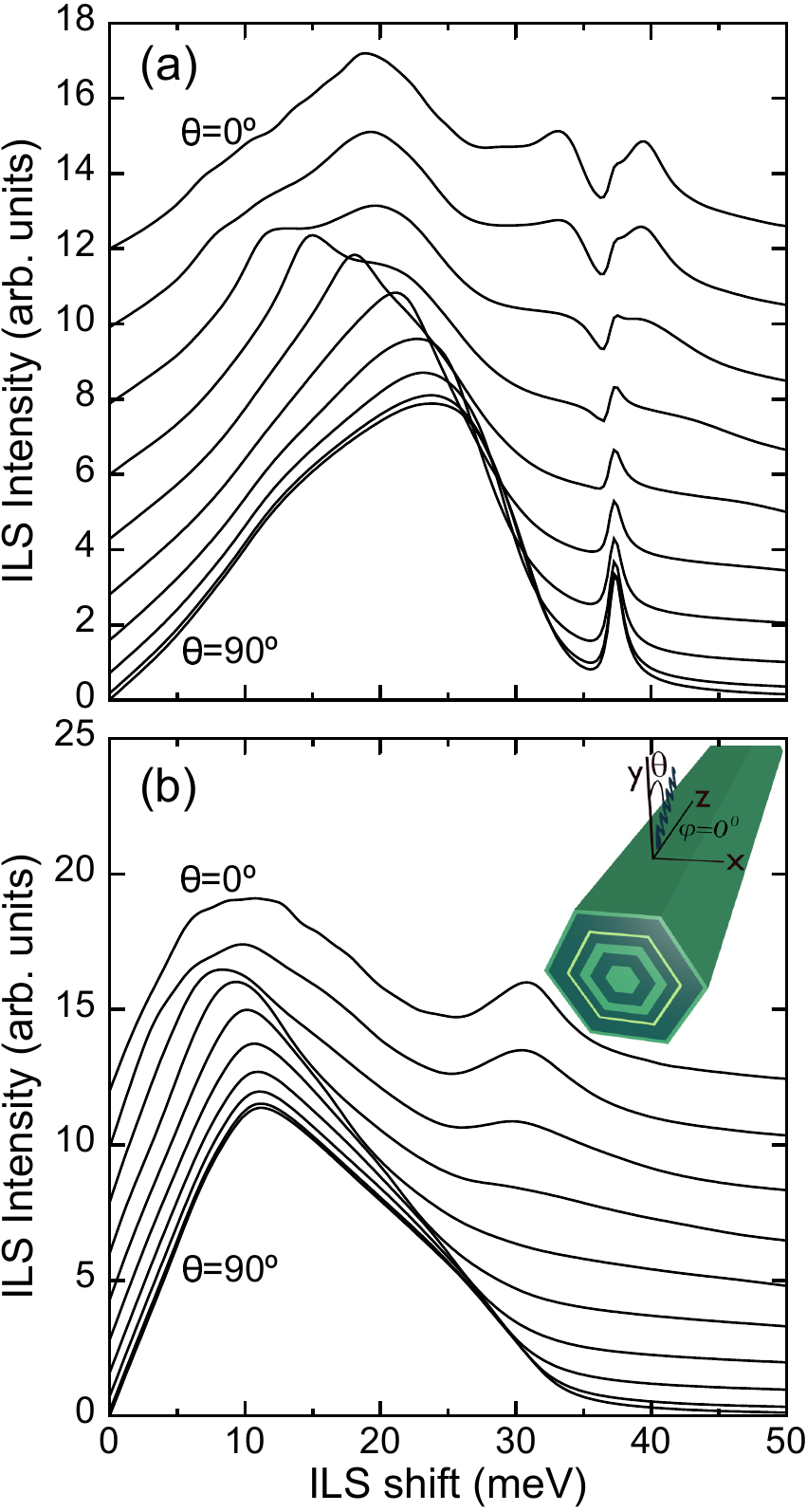}
\caption{CDE (a) and SDE (b) spectra dispersion over the $(y,z)$ plane. Calculated spectra with different photon orientations $\theta$ in steps of $10^{\circ}$ are shown. Each curve is shifted proportionally to the in-wire transferred momentum $q_z$. The top curves correspond to $q_z=0$, the bottom curves correspond to $q=\sim 1.03\,\mathrm{k_F}$. The inset in (b) illustrates the scattering geometry.}
\label{fig5}
\end{figure}

We now study the dispersion of the ILS spectra as the direction of the photons is tilted (always assuming a back-scattering geometry) in the longitudinal direction, along the CSNW axis. We indicate with $\mathrm{\theta}$ the angle between the direction of the photon beam and the $y$-axis, and with $\mathrm{\varphi}$ the angle between the projection of the beam on the $(x,z)$ plane with the $z$-axis (see inset in Fig.~\ref{fig5}(b)). We consider the dispersion over the $(y,z)$ plane ($\mathrm{\varphi}=0^{\circ}$) as the beam is tilted from normal to parallel to the CSNW top facet (i.e., from $\mathrm{\theta}= 0^{\circ}$ to $\mathrm{\theta}= 90^{\circ}$). 
In this configuration the momentum transferred along the invariant direction excites genuine intra-subband transitions within the 1D parabolic subbands. At large angles, moreover, SPE continua gain spectral weight and induce Landau damping, which in a hexagonal NW is a symmetry restrictive process.~\cite{RoyoPRB14} 

\begin{figure}[ht!]
\includegraphics[width=0.4\textwidth]{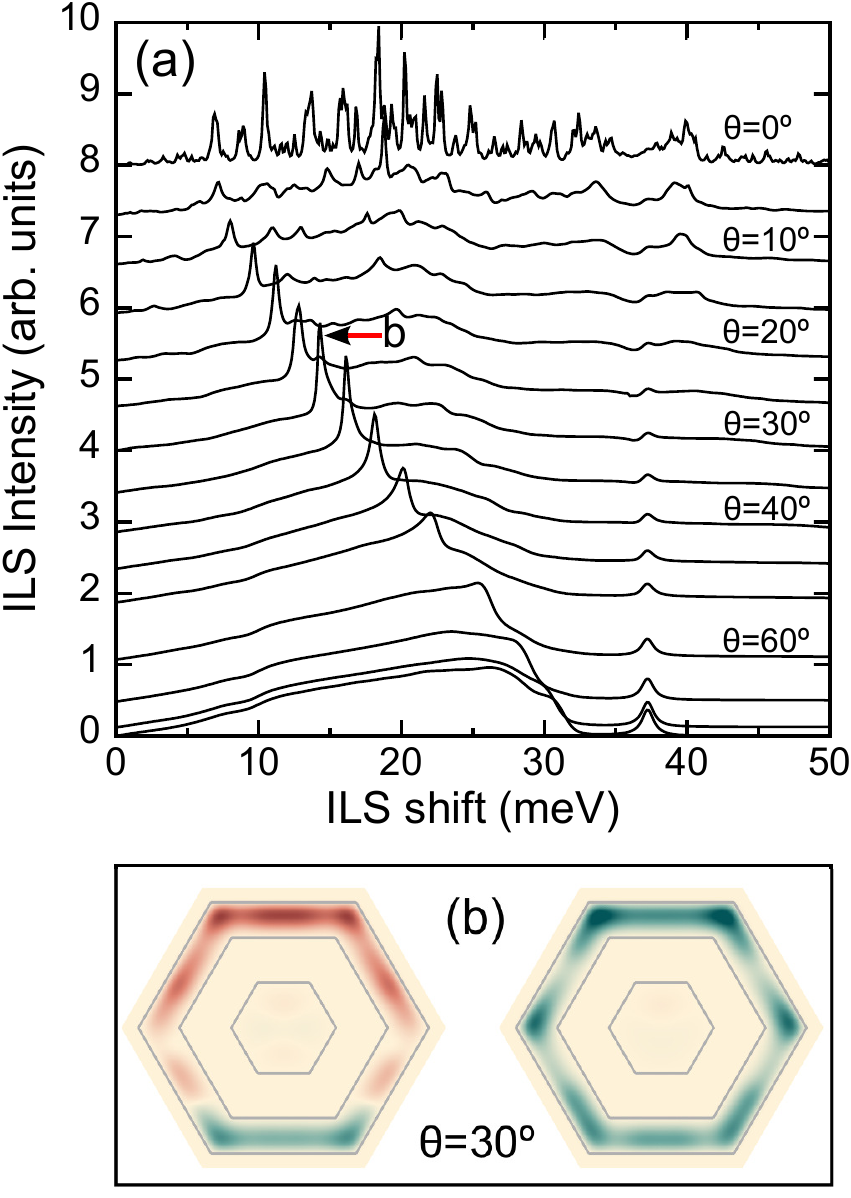}
\caption{(a) CDE spectra dispersion over the $(y,z)$ plane calculated using an electron damping of 0.1 meV. Curves corresponding to different photon orientations are shown for selected values of $\theta$ and have been shifted proportionally to the in-wire transferred momentum $q_z$. (b) IDD calculated at the resonant position of the long-lived CDE marked with an arrow in Fig. (a) for $\theta=30^{\circ}$.}
\label{fig6}
\end{figure}

The $(y,z)$ plane dispersion of the CDE and SDE spectra are respectively shown in Figs.~\ref{fig5} (a) and (b). Clearly, the spectra are strongly affected by the photon orientation. A main feature is the disappearance of the inter-subband resonances as the photons are tilted toward the longitudinal  direction. For SDE, e.g, the peak at $\sim 31$ meV disappears at $\mathrm{\theta} \gtrsim  30^{\circ}$. 
The inter-subband phonon-coupled peaks in the CDE spectrum also disappear, but a narrow peak appears at $\sim 39$ meV as a consequence of the coupling between the phonons and the persisting lower energy broad band which approaches the phonon resonances as $\mathrm{\theta}$ increases.

Vanishing of the inter-subband resonances with increasing transferred momentum in the invariant direction has been reported for planar systems both in simulations~\cite{HaiPRB98} and experiments.~\cite{SooryakumarPRB85,SchmellerPRB94}
This is due, on the one hand, to Landau damping of some collective modes inside SPE continua and, on the other hand, to the reduction of the momentum transferred normal to the QW plane. In our calculations this results from the ILS selection rules arising from the integrals in Eq.~(\ref{eq1}), which vanish for inter-subband transitions when $\mathbf{q}=0$ ($\theta=90^{\circ}$). Correspondingly, the 2D intra-subband excitations associated with azimuthal density fluctuations are also expected to weaken as $\mathrm{\theta}$ is increased since the momentum transferred along the azimuthal direction is reduced. In fact, the the 2D intra-subband CDE resonance at $\sim 19$ meV for $\theta=0^{\circ}$ (top curve in Fig.~\ref{fig5}(a)) reduces in intensity as $\theta$ increases. However, an additional peak starts to be observed at $\theta=30^{\circ}$ as a low-energy shoulder of the 2D intra-subband resonance. This new peak originates in 1D intra-subband excitations and it shows the characteristic strong blueshift with $q_z$ of an intra-subband plasmon.~\cite{RoyoPRB14} 

In order to better analyze this resonance, in Fig.~\ref{fig6}(a) we examine its dispersion in detail using a smaller electron damping (0.1 meV). The figure shows that the resonance is originated in a single collective mode which becomes the only one optically active for $\theta\geq20^{\circ}$. Therefore, contrary to the numerous 2D intra-subband and inter-subband modes showed in Fig.~\ref{fig4}, only a single intra-subband plasmon is observed in spite of the high number of occupied subbands. A similar result was reported in Ref.~\onlinecite{LinPRB93} for a zero-thickness cylindrical tubule. Specifically, when multiple 1D subbands are occupied, one among all possible intra-subband plasmons gets an increasingly larger spectral weight. This is the intra-subband plasmon corresponding to a coherent oscillation of all electrons along the longitudinal direction of system.
The IDDs calculated for the long-lived peak at $\theta=30^{\circ}$ (Figs.~\ref{fig6}(b)) confirms this assignment, since almost no density fluctuation is found over the CSNW cross section because the 1D intra-subband character of the mode induces density fluctuations in the longitudinal direction.

Finally, note that in the longitudinal limit ($\theta\geq60^{\circ}$ in Fig.~\ref{fig5}(a)) the ILS intensity is exclusively due to the intra-subband SPE continuum, and the long-lived intra-subband plasmon is Landau damped. This is a consequence of the effective dimensionality of the system induced by the large density of states along the facets, as pure 1D plasmons would not be Landau damped.~\cite{DasSarmaPRB96}

\subsection{In-plane dispersion  \label{Vargeom2}}

\begin{figure}[ht!]
\includegraphics[width=0.8\textwidth]{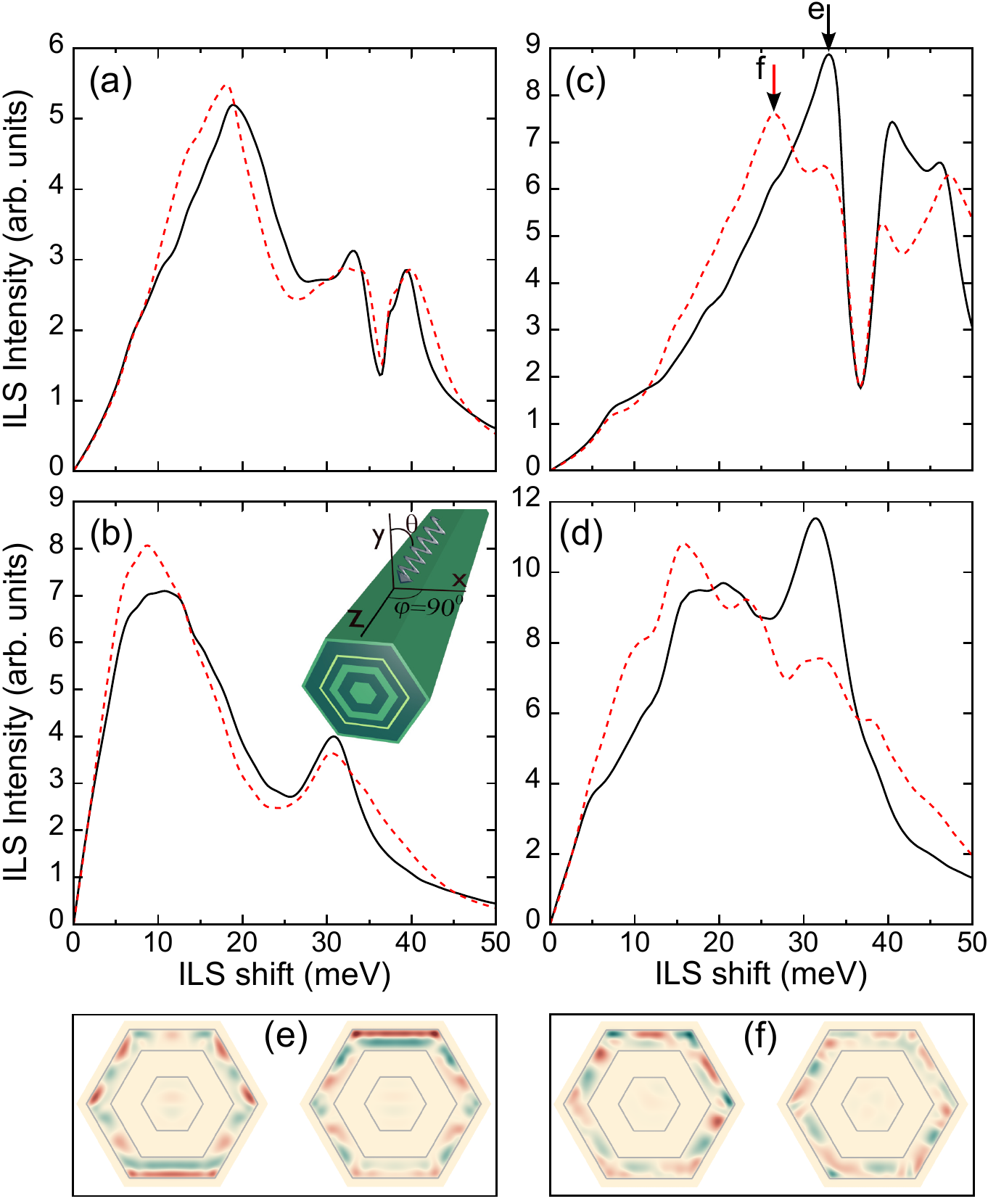}
\caption{(a) CDE and (b) SDE spectra for a sample with doping density $2\times10^{18}\, \mathrm{cm^{-3}}$. Black-solid (red-dashed) lines correspond to $\varphi=90^{\circ}$, $\theta=0^{\circ}$ ($\theta=30^{\circ}$). The inset illustrates the scattering geometry. (c) CDE and (d) SDE spectra with doping density $2.7\times10^{18}\, \mathrm{cm^{-3}}$. (e) and (f) are IDDs calculated at the CDE energies marked with arrows in panel (c).}
\label{fig7}
\end{figure}

We next discuss the dispersion when the photon beam is tilted in the plane of the cross-section of the CSNW, that is the $(x,y)$ plane (see Fig.~\ref{fig7}(b) inset). To estimate the expected anisotropy, we compare the spectra calculated at normal incidence ($\theta=0^{\circ}$) and with photons propagating parallel to a maximal diameter ($\theta=30^{\circ}$). ILS selection rules were discussed in Ref.~\onlinecite{RoyoPRB14} for these scattering configurations in a model coQW system with few occupied subbands.~\cite{RoyoPRB14} While some of the excitations are excited in both configurations, some other excitations are selectively totally or strongly suppressed in either configurations. Similar arguments apply to the present study. However, in a realistic experimental situation, which we simulate here using a large electron gas density and electron damping in the meV range, it would not be possible to single out specific excitations, due to the large density of states of the coQW. Nevertheless, these selection rules are at the origin of the results discussed below.

The CDE and SDE spectra for the sample studied in the previous sections are shown in Figs.~\ref{fig7} (a) and (b). Tilting the photon beam over the $(x,y)$ plane slightly reshapes the peaks but otherwise it has a small effect on the spectra. In general, at $\theta=30^{\circ}$ the intensity of the 2D intra-subband peaks is increased, while inter-subband peaks are weakened, in comparison with the peaks at $\theta=0^{\circ}$. This is originated in the fact that for $\theta=0^{\circ}$ more momentum is transfered in the radial than in the azimuthal direction, thus favoring inter-subband excitations, whereas the opposite holds for $\theta=30^{\circ}$.

In Figs.~\ref{fig7}(c) and (d) we show the spectra for a larger doping density, $2.7\times10^{18}\, \mathrm{cm^{-3}}$, with a total free electron density $\sim3.11\times10^{7}\, \mathrm{cm^{-1}}$.\footnote{We have used a basis set of 130 subbands to obtain converged spectra at this doping density.} 
Compared with the spectra for the lower density, all peaks appear blueshifted due to the larger subband splitting.
For CDEs, however, this is also due to the dynamic many-electron Coulomb contributions typical of larger electron densities.~\cite{FunkNL13} 
Moreover, the 2D intra-subband and the inter-subband peaks partially overlap due to the larger blueshift of the former. In this  larger carrier density regime, the spectra are strongly anisotropic in the $(x,y)$ plane. For instance, in the SDE spectrum for $\theta=0^{\circ}$ the inter-subband peak at $\sim 31$ is stronger than the 2D intra-subband resonance at $\sim 16$ meV, while the opposite is true at $\theta=30^{\circ} $ Similar, but less pronounced, reshaping is also predicted for the CDE spectrum. 
Indeed, the IDD calculated at the energy of the most intense CDE peak for $\theta=0^{\circ}$ (Fig.~\ref{fig7}(e)) shows mainly a radial density fluctuation in the top and bottom facets, while the most intense peak for $\theta=30^{\circ}$ (Fig.~\ref{fig7}(f)) is rather originated in azimuthal density fluctuations. 

\section{Summary and conclusions \label{conc}}

CSNWs are quite complex systems from the electronic point of view. On the one hand, due to the tubular shape, the electron gas bridges between a q2D and a q1D system. On the other hand, the prismatic cross-section induces an inhomogeneous distribution of carriers, forming q1D and q2D channels at the corners and facets of the heterostructure, respectively. Furthermore, in ILS experiments the photon field inevitably impinges with different directions on the different facets of the embedded coQW at the same time, exciting different types of excitations simultaneously. Therefore, ILS spectra in CSNWs are considerably more complex than in planar heterostructures. At the same time, ILS cross-section is an extremely informative probe of the nature of the electronic states, particularly if dispersion is measured in appropriate directions, as we have discussed in this paper. 

To show the potential of ILS experiments in CSNWs, we have predicted the spectra by CDEs and SDEs at experimentally relevant regimes. The real-space 3D multi-subband DFT and TDDFT formalism used here proved of predictive quality in a previous theoretical-experimental study which has been conducted at normal incidence.~\cite{FunkNL13} Here, we have generalized this approach to predict the dispersion of the ILS resonances with the photon field rotated in different directions. We have shown that the ILS spectra hold information on three different type of collective excitations which form in CSNWs, namely, 2D-like intra-subband, inter-subband and intra-subband excitations. These are associated with density fluctuations in the azimuthal, radial and longitudinal directions, respectively, and can be singled out in ILS from their spectral position and bandwidth, and from their specific dispersion in properly designed experiments. 

In particular, i) the 2D intra-subband and inter-subband peaks vanish as the photon field is rotated toward the CSNW axis; ii) in this regime a single long-lived intra-subband plasmon can be observed; iii) the ILS spectra are anisotropic with respect to the azimuthal angle of the photon field, the degree of anisotropy being larger for higher electron densities. 

\begin{acknowledgments}
We acknowledge partial financial support from Universitat Jaume I - Project P1.1B2014-24, from Generalitat Valenciana Vali+d Grant (MR) APOSTD/2013/052, from European Union’s 7th framework programme - Marie Curie ITN INDEX under grant agreement n° 289968, and from University of Modena and Reggio emilia, through Grant ``Nano- and emerging materials and systems for sustainable technologies''.
\end{acknowledgments}

\bibliography{Elementary_excitations}

\end{document}